\documentstyle[12pt,psfig,aasms4]{article}
\tightenlines

\newcommand{\op}{Ly$\alpha$\ }
\newcommand{\no}{log$N({\rm HI})$\ }

\begin{document}


\title{Temperature and Kinematics of CIV Absorption Systems}

\author{M. Rauch\altaffilmark{1,2}, W.L.W. Sargent\altaffilmark{1}, D.S. Womble\altaffilmark{1,2}, T.A. Barlow\altaffilmark{1}}
\altaffiltext{1}{Astronomy Dept., 105-24 California Institute of Technology, Pasadena 91125}
\altaffiltext{2}{Hubble Fellow}
\footnotetext{The observations were made at the W.M. Keck Observatory
which is operated as a scientific partnership between the California
Institute of Technology and the University of California; it was made
possible by the generous support of the W.M. Keck Foundation.}

\vskip 2.0cm

{Subject Headings:  Galaxies: formation, evolution ---
intergalactic medium --- quasars: absorption lines} 

\vskip 2.0cm
\centerline{\it submitted to the ApJ Letters, March 16, 1996
(in press).}
\vskip 3.0cm

\begin{abstract}
We use Keck HIRES spectra of three intermediate redshift QSOs to study
the physical state and kinematics of the individual components of CIV
selected heavy element absorption systems. Fewer than 8\% of all CIV
lines with column densities greater than 10$^{12.5}$ cm$^{-2}$ have
Doppler parameters b$<$ 6 kms$^{-1}$.  A formal decomposition into
thermal and non-thermal motion using the simultaneous presence of SiIV
gives a mean {\it thermal} Doppler parameter $b_{therm}(CIV)$=7.2
kms$^{-1}$, corresponding to a temperature of 3.8$\times 10^4 K$
although temperatures possibly in excess of 3$\times 10^5 K$ occur
occasionally.  We also find tentative evidence for a mild increase of
temperature with HI column density.  Non-thermal motions within
components are typically small ($<$10 kms$^{-1}$) for most systems,
indicative of a quiescent environment.  The two-point correlation
function (TPCF) of CIV systems on scales up to 500 kms$^{-1}$ 
suggests that there is more than one source of velocity
dispersion. The shape of the TPCF can be understood if the CIV systems
are caused by ensembles of objects with the kinematics of dwarf
galaxies on a small scale, while following the Hubble flow on a larger
scale.  Individual high redshift CIV components may be the building
blocks of future normal galaxies in a hierarchical structure formation
scenario.

\end{abstract}

\section{Introduction}
 
The CIV $\lambda\lambda$1548,1550 $\AA\ $ doublet is the most common
heavy element absorption feature found in the spectra of high redshift
quasars. Most CIV absorption systems must arise in
highly ionized (e.g. Bergeron \& Stasinska 1986) and strongly clustered
gas clouds (e.g. Sargent, Boksenberg \& Steidel 1988). 
Recent Keck observations (Cowie et al. 1995, Tytler et al. 1995) have
found CIV absorption to be common for Lyman $\alpha$ forest clouds
down to N(HI) $\sim 10^{14}$cm$^{-2}$. Thus CIV systems
trace the gas over the same column density range as the gaseous structures
apparently dominating the baryonic contents of the universe (at $z>2$)\  (Rauch \& Haehnelt 1995, Miralda-Escud\'e et al. 1996). 
These systems then may tells us about
the physical conditions in the main reservoir of matter, prior to incorporation   into virialized galaxies. 
The precise dynamical state
of the individual CIV components has remained elusive.  York et al.  (1984)
established upper limits on the temperature of the gas, ruling out
collisional ionization as the dominant equilibrium process.  Cowie et
al. (1995) found CIV and
corresponding HI Lyman $\alpha$ lines to be consistent with clusters of
components of photo-heated gas and a small internal velocity dispersion
of 18 km/s.  To date the existing measurements of the temperature have
strictly speaking been upper limits. In the present paper we revisit
the  thermal and kinematic properties of CIV systems, 
using high resolution spectroscopy on two ions with
different mass to disentangle the contributions of temperature and bulk
motion to the line profiles.  In particular we ask whether there is a
lower limit to the temperature and whether a possible range of temperatures
indicates heating sources other than photoionization. We also consider
the large scale motion of entire CIV complexes and try to model the
velocity structure with a minimum number of assumptions.

\section{The Data}

We have observed the spectra of three QSOs, 1225+315, 1107+487, and
1422+231  with the high resolution spectrograph of the Keck telescope. The exposure times were 10500, 26000, and 25100
seconds, respectively, resulting in typical signal-to-noise ratios of
20-35 (1225+315), 30-45 (1107+487), and 55-85 (1422+231) per 0.04 $\AA\ $ pixel. The observation of
Q1225+315 was done with a 0.57 arcsec wide slit to achieve very high
resolution of 5.3 kms$^{-1}$ (FWHM; as measured  from a ThAr
spectrum) for the final linearized and summed data, with the explicit
purpose of trying to resolve the thermal width of the CIV doublet. The spectra of the other two objects were obtained 
with a 0.86 arcsec
slit (FWHM 6.6 km$^{-1}$).  The
data were reduced as described in Barlow \& Sargent (1996, in prep.). A continuum was
fitted with spline functions, and the line profiles where fitted with
the Voigt profile package VPFIT (e.g. Carswell et al. 1991) 
The complete data sets will be published elsewhere together with the
analysis of the \op forest regions. 

\section{Analysis of the absorption line profiles}

Figure 1 (top) shows the Doppler parameter - column density distribution (b -
log N(CIV)) for all CIV components in the spectrum of Q1225+317 (the
highest resolution data set) longward of \op emission and with column
densities above logN(CIV)=12.  The dotted line indicates the lowest
Doppler parameter resolvable in any region of the spectrum, 3.18
kms$^{-1}$ (5.3 kms$^{-1}$ FWHM).  Wherever possible
these parameters were obtained from fits to both lines of the CIV
doublet. Note
that the large majority of CIV lines are resolved. Given the size of the error bars, quite a number of
points should have scattered below the dotted line if there were many
lines close to or narrower than the resolution limit.  
The bottom panel shows the same distribution for the full sample of all 3 QSOs.
Fig. 2 gives a Doppler parameter histogram for the full
sample. Mean and median Doppler parameters are 9.3$\pm$ 0.7
kms$^{-1}$, and 10.6 kms$^{-1}$, respectively. Only 15 \% of the lines have
b values below 6 kms$^{-1}$; for those lines with logN
$>$12.5 the fraction is lower at 8 \%. 81\% of all
lines have b $<$ 16 kms$^{-1}$.

To separate the contributions of thermal and non-thermal motion
("bulk motion", or "turbulence")  we use the simultaneous
presence of the CIV and SiIV ion, and we assume that the bulk
motion can be represented by a Gaussian (our main conclusions regarding
the temperature range are not strongly dependent on the exact
method of deconvolution).  Then we can write
\begin{eqnarray} b_i^2 = b_{b}^2 + \frac{2kT}{m_i},
(i=1,2), 
\end{eqnarray}
where the index i denotes the ion used, in our case CIV (1) and SiIV
(2).  We solve these equations for T and $b_b$.  This procedure is
based on the assumption that CIV and SiIV have the same temperature and bulk
motion in each cloud, inspite of their rather different
ionization potentials.  This assumption need not be universally true,
but it holds for a pressure confined
static gas cloud; hydrodynamic simulations (Haehnelt, Steinmetz \&
Rauch 1996) indicate that it is also approximately valid for gravitationally
collapsed clouds, because the spatial extent of the regions dominating
the line formation is determined by the spatial scale of highest peaks in the total gas
density, even for higher ions, and the ionization fraction is only of
secondary importance.

We solved the equations (1) for a
sample of 79 of the 208 CIV absorption systems for which both CIV
and SiIV were available. We have fitted the absorption complexes requiring the SiIV components to have the same redshifts as CIV.  
We have retained in our analysis only the CIV and SiIV absorption line pairs  with fractional errors
of less than 20 \% in their Doppler parameters in order to eliminate   
ill-constrained cases caused by run-away profile fits, noisy weak lines, and severe blending.  Five cases where the best fit Doppler
parameter of SiIV, b(SiIV), ended up larger than b(CIV) were excluded
from the analysis as being unphysical.  For the surviving  sample of 26 CIV/SiIV
pairs we computed the thermal Doppler parameter of the CIV component
from 
\begin{eqnarray} b^2_{therm_{CIV}} = \frac{b^2_{CIV} -
b^2_{SiIV}}{1-\frac{m_C}{m_{Si}}}. \end{eqnarray} 
The sample
$b_{therm}(CIV)$ versus total Doppler parameter $b(CIV)$ is plotted
in figure 3.  The solid lines are the loci of constant bulk motion,
beginning with zero, i.e., pure thermal motion ($b_{therm}(CIV)=b(CIV)$)
on the upper left and proceeding to $b_b(CIV)$ = 20 kms$^{-1}$ in steps of 5
kms$^{-1}$.  The dashed line denotes equipartition between thermal and non-thermal energy.
The measured thermal
Doppler parameters tend to lie mostly between the zero and the 10
kms$^{-1}$ bulk motion line. The mean of the $\it thermal$ CIV $b$
parameter for this sample is $\overline{b_{therm}(CIV)} = 7.2 $
kms$^{-1}$, The corresponding value for the $\it total$ CIV Doppler
parameter is $\overline{b(CIV)} = 9.6 $ kms$^{-1}$, similar to that of
the full sample of CIV lines.  The medians are $b^{med}_{therm}(CIV) = 9.9$ kms$^{-1}$,
and for the total width $b^{med}(CIV) = 12.6$ kms$^{-1}$, somewhat higher than the means because of a tail in the
distribution towards large Doppler parameters with relatively larger
errors.  A "typical" value for the bulk motion contribution is then given by
\begin{eqnarray}
b_b(CIV) = \sqrt{\overline{b(CIV)}^2 - \overline{b_{therm}(CIV)}^2} = 6.3 \ kms^{-1}
\end{eqnarray}
It appears that this value is not representative
of the full range in b as it is dominated by the low b lines which have
smaller errors.   An apparent increase of
$b_b(CIV)$ with the total b(CIV) beyond 15 kms$^{-1}$ formally implies
that hotter gas also has more bulk motion.  However, some of the
broader lines may well be blends, so in theses cases we
may be overestimating the non-thermal component.
In any case, it is clear that the kinetic energy of the gas causing
these high ionization absorption systems is dominated by
thermal energy, and that the contribution from turbulence or other
(quasi-Gaussian) bulk motions are small.  Even if we remain sceptical
about the Gaussian assumption we can always take the total
Doppler parameters as {\it upper limits} to any contribution from bulk
motion.

The value $\overline{b_{therm}(CIV)}$ = 7.2 kms$^{-1}$ translates into
a temperature T $\approx$ 38000 K, or a thermal contribution to the \op
line width $b_{therm}(HI)$ of 25.0 kms$^{-1}$.  If, as we have seen,
above the thermal Doppler parameter occupies a relatively narrow range
some of the broadest HI \op lines must be due to unresolved clusters
(Rauch et al. 1992).  Therefore, the difference between the average
total Doppler parameter for the Lyman $\alpha$ forest and the smaller
thermal widths measured here may indeed often be one of small scale
velocity dispersion among several individual CIV components, as
suggested by Cowie et al. (1995).

If we take the minimum HI Doppler parameter ($\sim$ 20 kms$^{-1}$) of
the {\it lower} column density ($3\times10^{13} < N(HI) <
3\times10^{14}$) systems in the Hu et al. (1995) Ly$\alpha$ forest
sample to correspond to pure thermal broadening then the thermal HI b
parameter extrapolated from our {\it higher} column density (all lines
correspond to  $N(HI)>$ a few times $10^{14}$)  CIV sample indicates
a mild increase of b with N(HI), with b$_{therm}$(HI) rising
from $\sim$ 20 to 25 kms$^{-1}$ (T from 2.5-3.8$\times 10^4$ K), while
the column density increases by at least a factor 10. Taken together
with our finding of a range of temperatures  up to $3\times10^5$ K and
possibly beyond (fig. 3), this implies that not
all high column density, broader absorption lines are small
clusters of subunits with a uniform, low temperature.

\section{Understanding the kinematics of CIV systems}

The two point auto-correlation function (TPCF) of the full CIV sample
is given in figure 4, out to a pair splitting of 700 kms$^{-1}$.  The
amplitude of each bin of the TPCF has been normalized by the expected
number of counts per bin for a random distribution of clouds.

After a steep decline within the first 200 kms$^{-1}$ there is a
flattening in the TPCF out to about 400 kms$^{-1}$, before the
amplitude drops to the average value or even below.  Following
Petitjean and Bergeron (1990,1994, hereafter PB) we can model the shape
of the TPCF formally by fitting 3 Gaussians to the data, with velocity
dispersions $\sigma$=22, 136 and 300  kms$^{-1}$, respectively.  PB found a best fit for two Gaussians, with $\sigma$
values of 80 and 390 kms$^{-1}$ for a $<z>\sim$ 1 MgII sample, and
$\sigma$'s of 109 and 525 kms$^{-1}$ for a CIV sample with
$<z>$=2.65.  In our CIV data the third Gaussian was required to match
the narrow peak at zero velocity; such a detail may not have been obvious
in the PB sample.   The similarity between the MgII and CIV TPCFs is
quite striking, as PB noted, and this coincidence may be an
argument in favour of QSO absorbers being small
clouds each with high and low ionization zones subject to the same
kinematical conditions, rather than large onion shell structures.

The large velocity splittings in the observed TPCF may result
from  ordered large scale motion rather than isotropic random
velocity dispersion. Accordingly, we investigate a simple model, where a
set of identical "absorbers" occupies random positions
within an extended sheet, which in reality may represent galaxies
embedded in a large scale wall of
gas. The sheet is taken to
expand with the Hubble flow, and the individual
absorption components are assumed to have an internal isotropic Gaussian velocity distribution.  We have computed the TPCF by running randomly directed lines-of-sight through the centers of these slabs.
The result of one such calculation are shown overplotted as a
dotted line on the observed TPCF in figure 4. For this
realization, the sheets were 1.5 Mpc long and wide, and 60
kpc thick, and
the Hubble constant was 550 kms$^{-1}$ Mpc$^{-1}$ (at z=2.78). The
internal velocity dispersion among the components was  $\sigma$ =20
kms$^{-1}$, and the absorbers had hard boundaries with radii of 45 kpc. The
number density of absorbers {\it within the sheets} was 300 Mpc$^{-3}$,
with the overall normalization chosen
such as to match the observed TPCF.
Obviously, a simple model with a small internal velocity dispersion
coupled with an nonisotropic expansion of randomly oriented sheets of
clouds can match the measured TPCF quite well. This probably
does not rule out alternative models where the velocity
splitting is dynamical, e.g. with CIV systems orbiting
in large galactic halos ($v_c\sim$ 200 kms$^{-1}$), but it seems that
deep potential wells are {\it not required} to explain the TPCF.
If this picture of smallish CIV clumps in expanding, coherent large scale 
walls (surrounding voids) is correct, the {\it largest velocity splitting}
is then a measure of the product
of Hubble constant and coherence length of the walls, $\Delta v_{max}
\approx H(z) L(z)$.  This quantity increases with redshift only $\propto (1+z)^{\frac{1}{2}}$
($q_0$=0.5), so we should expect the temporal evolution of
the high velocity tail of the TPCF to be weak, consistent with
PB's observed small difference between the low z MgII and high z CIV correlation
functions.

\section{Conclusions}

The kinetic energy of each individual CIV component
is dominated by thermal motions with average temperature $\sim$ 38000
K, occasionally ranging up to a few hundred thousand degrees.  The 
lower limit to the CIV Doppler parameter and a likely positive
correlation between temperature and column density also point to gas heated by photoionization, perhaps compressed under the
influence of a shallow potential well.  However, individual CIV
components do not display a velocity broadening characteristic of the
full virial velocity of a massive galaxy but are at best compatible 
with the velocity dispersion in dwarf
spheroidals (e.g.  Hargreaves et al.  1994).  To produce line widths as
small as measured, the individual clouds must be rather quiescent and
could be confined quasi-statically by an isotropic outer pressure from
a hotter medium (e.g. Mo 1994). Alternatively, since the absorption
line optical depth is proportional to the gas density, the CIV lines
could arise in spatially small, collapsed regions (e.g., cooling gas
at rest behind an accretion shock), where the strong density gradients assure
that only the quiescent, narrow density peaks contribute noticeably to the line profile.

A simple model of an ensemble of absorbing objects with small
internal velocity dispersion, embedded in flattened large scale
"walls" expanding with the Hubble flow can explain the measured TPCF, including the narrow maximum at
the smallest splittings, and the tail out to 400km/s, which in this
model is due to chance alignments of our line of sight with the
expanding large scale structure. 

The large velocity spreads of many hundred km/s and the quiescent structure
of the individual components are consistent with the identification of CIV
systems with groups of protogalactic clumps ($M_{baryon}\sim 10^9 M\odot$) in
a hierarchical structure formation scenario (Haehnelt, Steinmetz \& Rauch 1996, Rauch et al. in prep.). If this conjecture
should turn out to be correct, high resolution observations of CIV
systems should be able to provide us with unprecedented insights
into gravitational galaxy formation at high redshift.

\acknowledgments
DSW and MR are grateful to NASA for support through grants
HF-1040.01-92A and HF-01075.01-94A from the Space Telescope Science
Institute. WLWS was supported by grant AST92-21365 from the National Science
Foundation.  We thank Martin Haehnelt for discussions, Bob Carswell
and John Webb for providing us with their profile
fitting software, and the W.M. Keck Observatory Staff for their help
with the observations.

\pagebreak
{\bf Figure captions:}

\bigskip

\figcaption[fig1_top.ps,fig1_bot.ps]{Above: Doppler parameter versus column density for CIV systems in the
line-of-sight to QSO 1225+315. Below: the same plot for the full sample of 208 CIV components.
The dotted line gives the minimum resolution attained in each sample.
Two points (probable run-away fits) at 41 and 57 km/s are off the edges.}

\figcaption[fig2.ps]{Doppler parameter histogram for the full sample}

\figcaption[fig3.ps]{thermal versus total CIV Doppler parameter for 26 systems selected
to have small relative errors in both b(CIV) and b(SiIV). The solid curves
are lines of constant non-thermal motion, the dashed line is the locus 
of equipartition between thermal and non-thermal energy.}

\figcaption[fig4.ps]{two-point correlation function (TPCF) for the full CIV sample, 
normalized by the expected number of pairs per bin for a random distribution
in redshift space. The dotted line is the TPCF
for the expanding-sheet model described in the text.}

\pagebreak
\newpage

\vspace{-2.cm}
\begin{figure}[htb]
\centerline{
\psfig{figure=civfig1_top.ps,width=10.cm,angle=-90.}
}
\vskip 0.cm
\centerline{}
\centerline{
\psfig{figure=civfig1_bot.ps,width=10.cm,angle=-90.}
}
\vskip 0.cm
\centerline{Figure 1}
\end{figure}
\newpage
\vspace{-2.cm}
\begin{figure}[htb]
\centerline{
\psfig{figure=civfig2.ps,width=10.cm,angle=-90.}
}
\vskip 1.cm
\centerline{Figure 2}
\end{figure}
\newpage

\begin{figure}[htb]
\centerline{
\psfig{figure=civfig3.ps,width=11.cm,angle=-90.}
}
\vskip 2.cm
\centerline{Figure 3}
\end{figure}
\newpage

\begin{figure}[htb]
\centerline{
\psfig{figure=civfig4.ps,width=10.cm,angle=-90.}
}
\vskip 1.cm
\centerline{Figure 4}
\end{figure}
\newpage

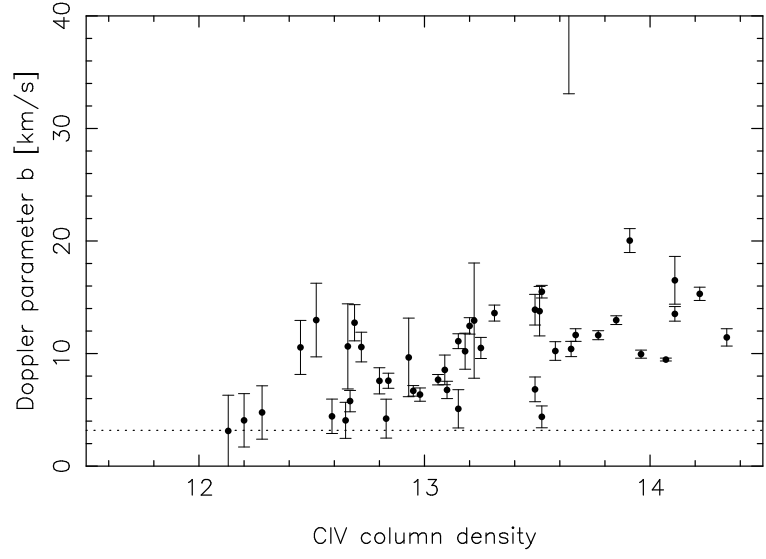
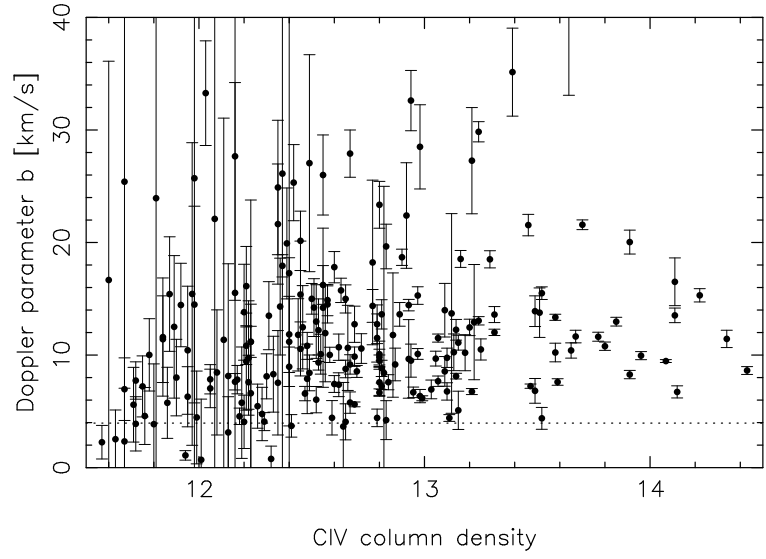
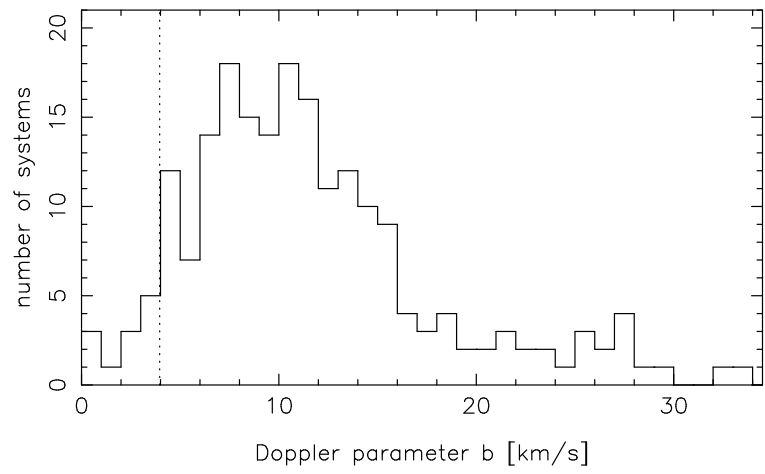
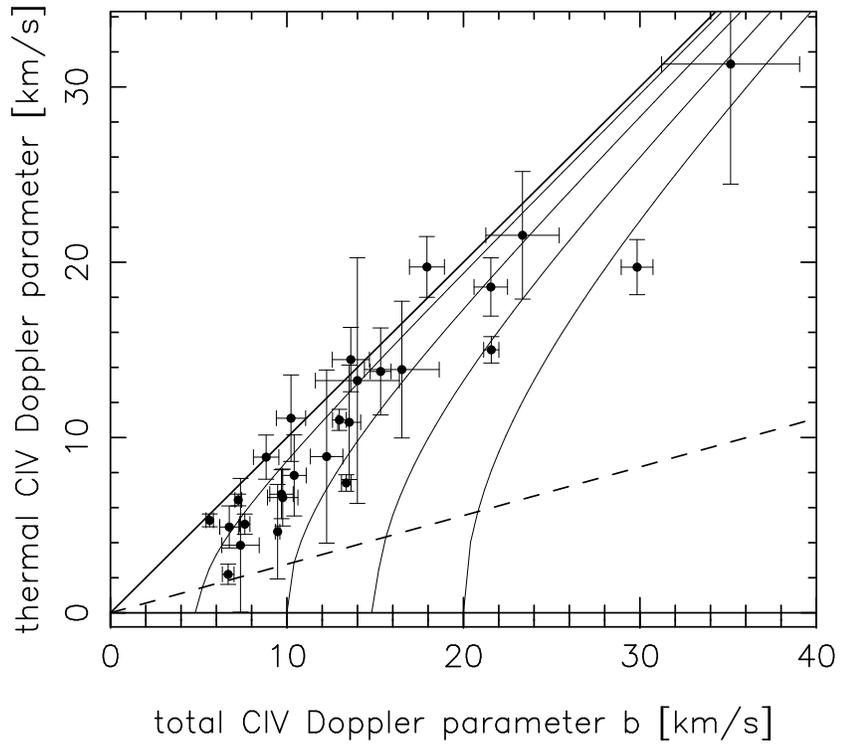
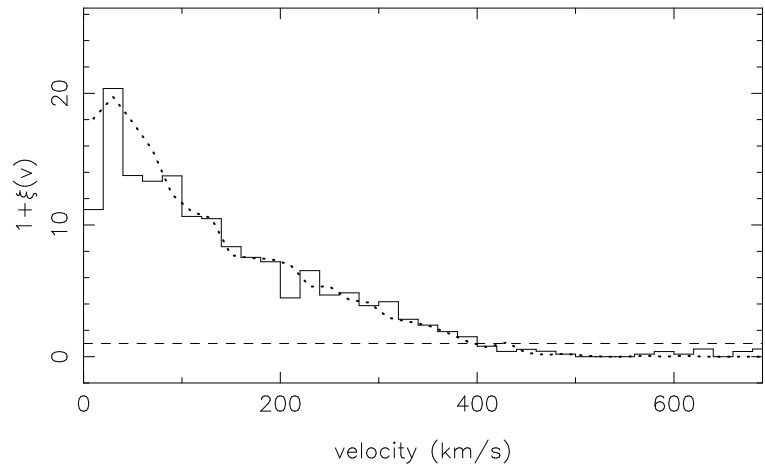
\end{document}